\documentstyle[aps,multicol,epsf]{revtex}

\begin{document}
\draft

\title{Microscopic Model for Granular Stratification and Segregation}
\author{Hern\'an A. Makse,$^{1,2}$ and Hans J. Herrmann$^{1,3}$}
 
\address{
$^1$P.M.M.H., Ecole Sup\'erieure de Physique et de Chimie Industielles, 10 rue
Vauquelin, 
75231 Paris Cedex 05, France\\
$^2$Center for Polymer Studies, Physics Dept., Boston
University, Boston, MA 02215 USA\\
$^3$Institute for Computer Applications 1, University of Stuttgart,
Pfaffenwaldring 27, 70569 Stuttgart, Germany}
\date{Europhys. Lett. {\bf 43}, 1-6 (1998)}
\maketitle
\begin{abstract}
We study  segregation and stratification 
of mixtures of grains differing in size, shape and material properties
poured in  two-dimensional silos using a microscopic lattice model for
surface flows of grains. 
The model incorporates the dissipation of energy in collisions between
rolling and static grains and an energy barrier describing
the geometrical asperities of the grains.
We study the phase diagram of the different morphologies 
predicted by the model as a function of the two
parameters. We find  regions of segregation and
stratification, in agreement with experimental finding, as well as a
region of total mixing.
\end{abstract}

\pacs{PACS Numbers: 05.40+j, 46.10+z, 64.75+g}

\begin{multicols}{2}

\narrowtext

When mixtures of grains
\cite{bagnold,borges,review1,review2,review3,review4,review5,review6} 
of different sizes 
are poured on a heap, a size segregation of the mixture is observed;
 the large grains
are 
more likely to be found near
the base, while the small grains are more likely to be near the top
\cite{segregation1,segregation2,segregation3,segregation4,segregation5,segregation6}.
If the grains differ not only in size but also in shape and roughness,
a spontaneous periodic pattern arises upon pouring the mixture 
in a two-dimensional cell.
When a mixture  of large-cubic grains and small-rounded grains
is poured in a vertical Hele-Shaw cell (two vertical slabs separated by
a gap of approximately 5 mm) the
mixture spontaneously stratifies into alternating layers of
small-rounded 
and large-cubic
grains  \cite{makse1}.
Otherwise, the mixture  only  segregates
when 
the large grains are more rounded  than the small grains  \cite{makse1}
with the large-rounded grains 
being found near the bottom of the cell.

The dynamical process leading to stratification was recently studied
numerically
and theoretically \cite{makse2},
using the set of continuum equations for surface flows of
granular mixtures developed in \cite{bouchaud1,bouchaud2,pgg,bdg}.
The physical quantities defined in this
phenomenological formalism 
 are to be understood as an average over a
certain coarse grain length on the surface of the sandpile (larger than
the size of the grains), where
hydrodynamic equations are valid, and any quantity defined below
this scale is not well-defined. 
Thus the relevant 
length scale appearing in this formalism
is that of
the
coarse grain scale ($\approx 5 d$) and  not of the grain size ($d$).
However
fluctuations may also occurs  at the level of the
grains, so that a microscopic description of collisions and transport
of grains may be needed to describe this situation.
In this paper, we study the dynamical segregation 
process in two-dimensional silos by
using a microscopic model of grain interactions.

We start by  defining the model for the case of a single species 
pile \cite{alonso,rimmele}, and then we discuss the
generalization to two types of grains differing in size and shape.
The microscopic model is defined on a square lattice.
Each grain has  width and height of one pixel.
Since the experiments are done by
pouring a fixed flux of grains,
we deposit $N$ grains at a given time step
at the top of the first column of the pile.
The grains start with a certain initial kinetic energy $e_0$, which 
will be lost 
in collisions with the static grains of the pile as the grains move down
the slope. 
Only one rolling grain is in contact with the pile surface, and therefore
interacts with the static grains of the pile. 
The remaining rolling grains are convected downward with unit velocity without
loosing their energy,
i.e., they move to the nearest-neighbor right column.
The loss of energy of the rolling grain interacting with the surface
pile is determined by the 
 restitution coefficient, $r$, which gives the loss of energy per unit
time. The interacting grain moves until its energy  is smaller than a
certain energy barrier $u$ and stops.
When the interacting rolling grain stops, one of the remaining rolling 
grains starts to interact with the surface pile
until it losses its energy and stops.
When all the $N$ rolling grains
stop, a new set of $N$ grains is 
dropped at the first column of the pile, and the
same rules are applied again.

The dynamics of a 
rolling grain with energy $e$ interacting with a static grain
located at height $h_i$ 
are defined as follows: 

\begin{itemize}

\item We test if the nearest neighbor 
position $i+1$ is energetically favorable, by first
calculating the energy test $e_{test}$, defined as the energy the grain would
have after moving to the new position $i+1$, according to

\begin{equation}
e_{test} = (e + \Delta h ) * r, 
\label{test}
\end{equation}
where $\Delta h = h_{i} - h_{i+1}$.

\item Then, the energy test is compared with the energy barrier $u$. If
$e_{test} > u$, the grain moves to the nearest neighbor position
$i+1$, and the energy of the grains is updated: $e = e_{test}$. Then the
procedure is applied to the interacting grain at column $i+1$, and so on.
If $e_{test} \le u$, the interacting grain stops at position $i$ and increases the height of the pile at $i$ by unity. The remaining rolling grains are always
convected downward to the position $i+1$.

\end{itemize}

Simulations and analytical calculations \cite{alonso}
show that the angle of repose of the pile is an increasing
function of the energy barrier $u$, and a decreasing function of
the restitution coefficient $r$.

\begin{figure}
\centerline{
\vbox{ \epsfxsize=5cm \epsfbox{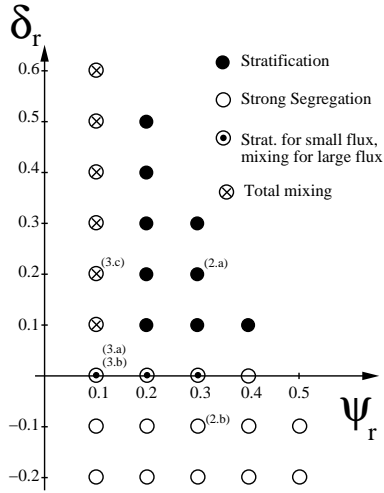} }        }
\vspace*{.5cm} 
\narrowtext
\caption{Phase diagram for surface flows of granular mixtures predicted
by the present model. In all simulations
we deposit two grains of type 1 and two
grains of type 2 ($N_1=N_2=2$), and we set $e_0 = 1$, and $\psi_u=0.3$.
For the region where $\delta_r>0$
we use $\delta_u=0.3$,  
for the region where
$\delta_r<0$
we use $\delta_u=-0.3$, and for the intermediate region of 
$\delta_r=0$ we use 
$\delta_u=0$. 
The letters indicate
the simulations
shown in Figs. 
\protect\ref{results} and \protect\ref{results-special}.}
\label{phase}
\end{figure}

Next we generalize the model to the case of two types of grains.
In \cite{makse2} stratification was reproduced using a discrete model 
defining different critical angles 
between the
grains.
The critical angle is defined as 
the maximum angle at which a rolling grain will be
converted into static grain.
This angle depends on the type of rolling
grain
and the type of static grain which is interacting with.
Thus, for two types of grains there are four different critical angles
also called generalized angles of repose $\theta_{\alpha\beta}$, with
$\alpha, \beta = 1, 2$.
Stratification of grains differing in size and shape
is the result of a
competition between size segregation and shape segregation \cite{makse3}. 
This
competition was incorporated in the models of \cite{makse2} at a macroscopic
level, by
considering certain relations between the generalized angles of repose.
The angle of repose of the pure species depends on the shape
of the grains: the rougher the grains the larger the angle of repose.
Thus, for mixtures of cubic grains (type 2) and rounded grains
(type 1) we have $\theta_{22}>\theta_{11}$.
On the other hand,
if the grains have different size, the cross-angles of
repose
$\theta_{\alpha \beta}$ are different. Since small grains roll down
on top of large grains easier than large grains on top of small grains,
this implies that $\theta_{12} > \theta_{21}$, if type 1
grains are smaller than type 2 grains.
Thus we arrive to the relations  for the case
of a granular mixture composed of small-rounded grains (type 1), and
large-cubic grains (type 2), which gives rise to stratification:

\begin{equation}
\label{thetast}
\theta_{21} < \theta_{11} < \theta_{22} < \theta_{12}.
\end{equation}

On the other hand, if the large grains are more rounded than the small
grains, one expects that $\theta_{22} < \theta_{11}$, and
this type of mixture results only in segregation and not in
stratification.
The 
relation between the angles of repose is then

\begin{equation}
\label{thetase}
\theta_{21} < \theta_{22} < \theta_{11} < \theta_{12},
\end{equation}
which is valid for small-cubic grains (type 1), and large rounded grains
(type 2).


\begin{figure}
\centerline{
\vspace*{.5cm} {\bf a} 
\epsfxsize=6.cm \epsfbox{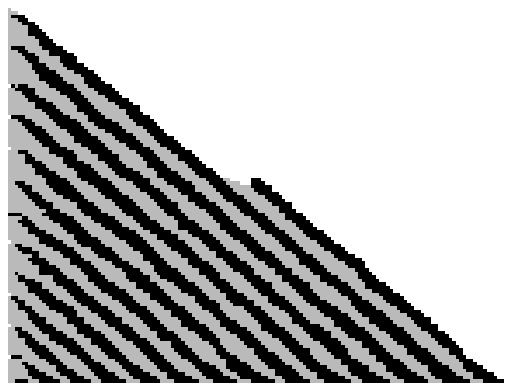}
} 
\centerline{
\vspace*{.5cm} 
{\bf b} \epsfxsize=6.cm \epsfbox{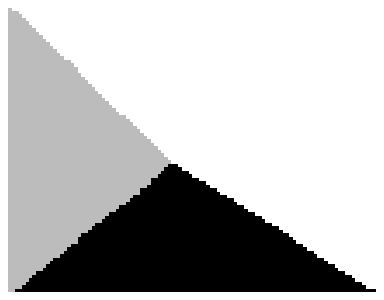} 
        }
\narrowtext
\caption{Different morphologies predicted by the present model.
{\bf a,} Stratification ($\delta_r = 0.2$, $\psi_r = 0.3$,
$\delta_u = 0.3$, $\psi_u = 0.3$,
$u_{21} = 0.3$, and 
$r_{12}=0.1$). Notice the ``kink'' 
formed by a pair of layers developed at the surface of the pile, 
similarly observed in \protect\cite{makse1,makse2}.
 {\bf b}, Strong segregation  
($\delta_r = -0.1$, $\psi_r = 0.3$,
$\delta_u = -0.3$, $\psi_u = 0.6$,
$u_{21} = 0.3$ and 
$r_{12}=0.1$).
Here the black grains are type 2, and grey grains are type 1.
See Fig. \protect\ref{phase} for the location of the
morphologies in the phase space.}
\label{results}
\end{figure}

Next, we 
generalize the model to the case of two types of grains $1$ and $2$
with
different size, shapes or  material properties.
We deposit $N_1$ and $N_2$ grains on the first column of the 
pile, and at a given time step, one rolling grain per species 
interacts with the
sandpile surface, and the remaining rolling grains move downward to  the
nearest-neighbor column.
We assume an overdamped situation where
the rolling grains which do not interact with the surface
achieve a constant convective velocity
due to the collisions with other rolling grains. 
We define the generalized
restitution coefficient and the generalized energy barrier
as
$r_{\alpha\beta}$, and $u_{\alpha\beta}$ with $\alpha, \beta = 1, 2$,
for
the four different possible collisions, i.e. $r_{12}$ is used in Eq.
 (\ref{test}) if a rolling grain of type $1$ collides with a static grains of type $2$.
Since the angle of repose of the pure species is a
monotonic 
increasing function of the energy barrier $u$, and a monotonic  decreasing 
function of the restitution coefficient $r$, 
we can translate the 
relations (\ref{thetast}) for stratification and (\ref{thetase}) for
segregation into relations for 	$r_{\alpha\beta}$ and
$u_{\alpha\beta}$.
Thus, we expect stratification for small-smooth grains (type 1), and
large-rough grains (type 2) when

\begin{eqnarray}
u_{21} < u_{11} < u_{22} < u_{12}\\
r_{12} < r_{22} < r_{11} < r_{21},
\end{eqnarray}
and, we expect segregation for a mixture of small-rough (type 1) grains, and
large-smooth (type 2) grains when

\begin{eqnarray}
u_{21} < u_{22} < u_{11} < u_{12}\\
r_{12} < r_{11} < r_{22} < r_{21}.
\end{eqnarray}

\begin{figure}
\centerline{
\vspace*{.5cm}
\hbox{ {\bf a}
\epsfxsize=4.5cm \epsfbox{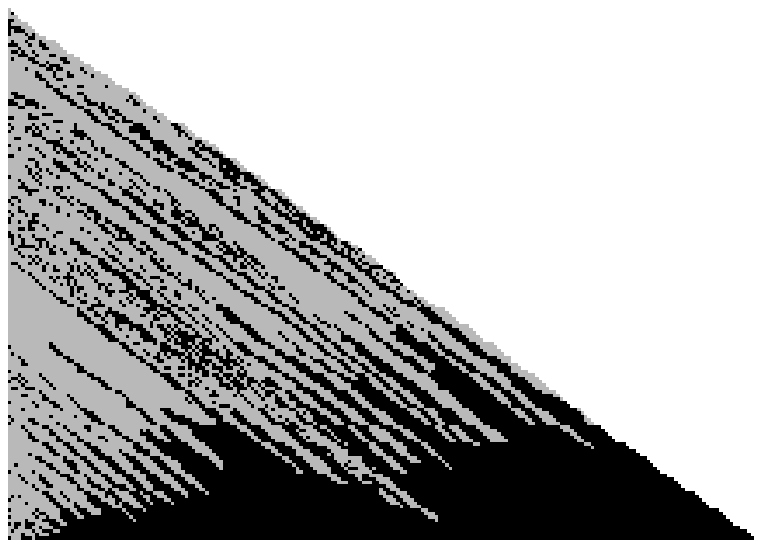} 
{\bf b} \epsfxsize=4.5cm \epsfbox{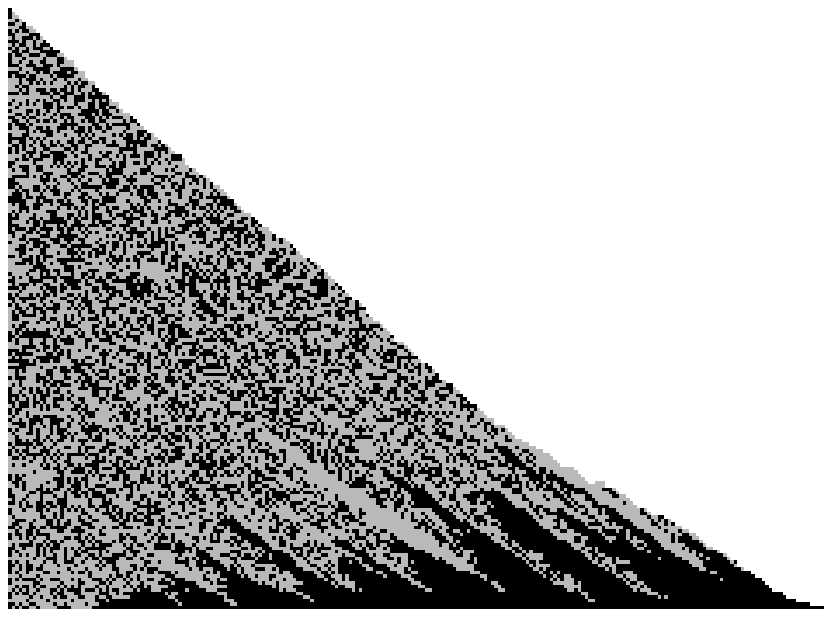} 
     }
        }
\centerline{
\vspace*{.5cm}
\hbox{ {\bf c} \epsfxsize=4.5cm \epsfbox{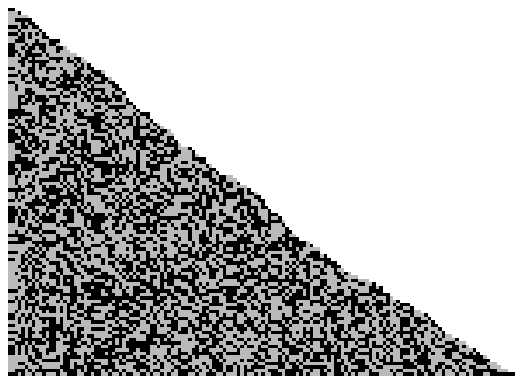} 
     }
        }
\narrowtext
\caption{Morphologies predicted by the present model when
the  grains are only slightly different.
{\bf a},
When $\delta_r = 0$ and $\psi_r\le 0.3$
 ($u_{21} = 0.1$, $r_{12}=0.4$) 
we observe weak stratification with thin and irregular layers
as observed in \protect\cite{baxter}
for small flux rate $N_1=N_2$=2,
and {\bf b}, 
mixing in almost all the pile
plus weak segregation and some stratification
with grains type 2 at the bottom of the pile 
for large flux rate $N_1=N_2$=16.
{\bf c}, 
When $\delta_r > 0$ and  $\psi_r = 0.1$ 
($u_{21} = 0.3$, $r_{12}=0.1$)
we observe the 
total mixing of the species. See Fig. 
\protect\ref{phase} for the location of the
morphologies in the phase space.}   
\label{results-special}
\end{figure}

We corroborate these predictions  by 
investigating the different morphologies predicted by the model
for the different internal parameters.
Since the general model has eight parameter, we reduce the number of them to be
able to investigate the resulting phase diagram.
We
assume some relations between the parameters and
define \cite{makse2} 
\begin{eqnarray}
\psi_u \equiv u_{11} - u_{21} =  u_{12} - u_{22}\\ 
\psi_r \equiv r_{11} - r_{21} =  r_{12} - r_{22},
\end{eqnarray}
and

\begin{eqnarray}
\delta_u \equiv u_{22} - u_{11} \\
\delta_r \equiv r_{11} - r_{22}.
\end{eqnarray}
Further, we  assume  some values for $\delta_u$ and $\psi_u$ (see Fig. 
\ref{phase}).
We notice that $\psi_u$ describes the difference in size of the grains, 
$\delta_u$ and $\delta_r$ are determined by the different shapes of the
grains, and the material properties and asperities
are described by
$\psi_r$ and $\delta_r$.
If grains 2 are
rougher that grains 1, we have  $\delta_r>0$, 
and if grains 2 are more smoother
than grains 1, we have $\delta_r <0$. 

Figure \ref{phase} shows the resulting phase diagram and
Figs. \ref{results} and \ref{results-special}
show the resulting morphologies. We find a region of
stratification when $\delta_r > 0$ (Fig. \ref{results}a),
 and a region of strong segregation 
for $\delta_r < 0$ (Fig. \ref{results}b),
as found in the experiments performed in
\cite{makse1,yan,kakalios}.

\begin{figure}
\centerline{
\vbox{ \epsfxsize=6cm \epsfbox{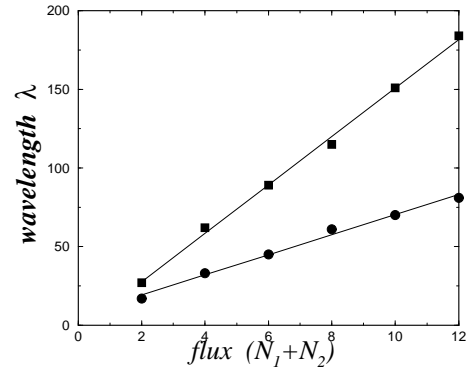} }
       }
\narrowtext
\caption{Wavelength of the layers (measured in pixel units)
as a function of the flux of incoming
grains ($N_1+N_2$)
for two different sets of parameters in the region of
stratification. 
We find a linear dependence in agreement with
conservation arguments \protect\cite{makse2}, 
the coefficient of the linear relation depends on the internal parameters.
The circles correspond to $u_{21} = 0.1$,
$u_{11} = 0.2$,
$u_{22} = 0.6$,
$u_{12} = 0.9$, while
the squares correspond to 
$u_{21} = 0.5$,
$u_{11} = 0.9$,
$u_{22} = 1.5$,
$u_{12} = 1.9$.
We take all the 
restitution coefficients 
equal to $0.2$, and we use an equal volume mixture ($N_1 = N_2$).
}
\label{lambda}
\end{figure}

When 
some of the properties of the grains are very similar
we find the additional morphologies
shown in Fig. \ref{results-special}. When $\delta_r = 0$ and $\psi_r \le 0.3$
we observe a weak stratification pattern
(Fig. \ref{results-special}a) but only for 
small flux rate $N_1=N_2=2$. The resulting layers are very thin
and irregular,
as can be seen in Fig.  \ref{results-special}a. The negligible
difference in grain properties gives rise to identical 
angles of repose of the pure species, so that the kink,
which is observed to give rise to the layers 
by stopping the rolling grains\cite{makse2}, is very small.
As a consequence, when we increase the flux of grains, the 
small kink is not able to stop the arriving rolling grains; 
the grains
ride over the kink so that not segregation at the kink is observed. 
Therefore, for these particular parameters,
the stratification pattern dissapears upon increasing the flux.
In Fig. \ref{results-special}b we show the results of our simulations
where the same calculations of  
Fig. \ref{results-special}a are done, but for a larger flux of grains
 $N_1=N_2=16$.
We see that at the 2/3 upper part of the pile
the grains are mixed, and that at the 1/3 lower part of the pile
there is some reminiscence of stratification plus the segregation
of the large grains at the bottom.
In general we find that when  
\( N_1=N_2  \stackrel{>}{\sim} 8\),
 the stratification disappears for
these kind of parameters, a prediction that was recently confirmed by
experiments \cite{baxter}.
Finally, when $\delta_r>0$ and $\psi_r = 0.1$ we observe the total mixing of the species for any value of the flux of grains (Fig. \ref{results-special}c).
As in the previous case, the mixing might be due to the existence of 
a weak kink that is not able to stop and segregate the grains.

We also study the wavelength $\lambda$ of the layers as a function of
the flux of grains (i.e., as a function of the total number $N_1+N_2$ of
deposited grains per unit time). We find a linear increase of $\lambda$
(Fig. \ref{lambda}) as a function of the flux as is expected from a
conservation argument \cite{makse1,makse2,yan}:

\begin{equation}
\lambda \propto N_1+N_2.
\end{equation}
We find that 
the coefficient of the linear relation depends on the internal parameters
of the model (Fig. \ref{lambda}). In general, we find that the larger
the difference in energy barrier of the two species, the larger the wavelength
of the layers, since the kink becomes steepest as the difference in angle of 
repose of the pure species increases.

ACKNOWLEDGEMENTS.
We thank J.J. Alonso, T. Boutreux, S. Luding, M. Rimmele, and H. E. Stanley
for stimulating discussions. H.A.M. acknowledges BP and CPS for
financial support.

\end{multicols}

\end{document}